# Topologically protected vortex transport via chiral-symmetric disclination

Zhichan Hu[1†], Domenico Bongiovanni[1,2†], Ziteng Wang[1†], Xiangdong Wang[1], Daohong Song[1,3], Jingjun Xu[1], Roberto Morandotti[2], Hrvoje Buljan[1,4*], and Zhigang Chen[1,3*]

[1]*The MOE Key Laboratory of Weak-Light Nonlinear Photonics, TEDA Applied Physics Institute and School of Physics, Nankai University, Tianjin 300457, China*
[2]*INRS-EMT, 1650 Blvd. Lionel-Boulet, Varennes, Quebec J3X 1S2, Canada*
[3]*Collaborative Innovation Center of Extreme Optics, Shanxi University, Taiyuan, Shanxi 030006, China*
[4]*Department of Physics, Faculty of Science, University of Zagreb, Bijenička c. 32, 10000 Zagreb, Croatia*
[†]*These authors contributed equally to this work*
*e-mail: hbuljan@phy.hr , zgchen@nankai.edu.cn

**Vortex phenomena are ubiquitous in nature, from vortices of quantum particles and living cells[1-7], to whirlpools, tornados, and spiral galaxies. Yet, effective control of vortex transport from one place to another at any scale has thus far remained a challenging goal. Here, by use of topological disclination[8,9], we demonstrate a scheme to confine and guide vortices of arbitrary high-order charges[10,11]. Such guidance demands a "double" topological protection: a nontrivial winding in momentum space due to chiral symmetry[12,13] and a nontrivial winding in real space arising from collective complex coupling between vortex modes. We unveil a vorticity-coordinated rotational symmetry, which sets up a universal relation between the topological charge of a guided vortex and the order of rotational symmetry of the disclination structure. As an example, we construct a $C_3$-symmetry photonic lattice with a single-core disclination, thereby achieving robust transport of an optical vortex with preserved orbital angular momentum (OAM) that corresponds solely to one excited vortex mode pinned at zero energy. Our work reveals a fundamental interplay of vorticity, disclination, and higher-order topological phases[14-16], applicable broadly to different fields and promising in particular for OAM-based photonic applications that require vortex guides, fibers[17,18], and lasers[19].**

All waves in nature can produce vortices[20], typically characterized by a flux circulation that gives rise to orbital angular momentum (OAM). Apart from the well-known vortices in classical fluids and optical fields[10,11], quantum vortices have been observed in superconductors, superfluids, as well as Bose-Einstein condensates[1,3]. Vortex beams carrying OAM have also been realized with photons[2], electrons[4], neutrons[5], and more recently non-elementary particles of atoms and molecules[6], encouraging many fundamental studies and unconventional applications.

Like all localized waves, however, vortices of any field tend to spread and deform during evolution. Confining a vortex flow is important in diverse areas of science and technology. For instance, in a cardiovascular channel, control and suppression of a vortex flow can lead to drag reduction. In aerodynamics, effective control of vortex generation and transport is essential for inhibiting turbulent boundary layer separation. In a thermonuclear fusion reactor, a zonal vortex flow can serve as a transport barrier to suppress heat losses through tokamak walls. In superconductors, individual vortex manipulation is pertinent for fundamental investigations and the development of superconducting devices[21]. Perhaps, the best-known examples are in optics: optical vortex beams (OVBs) and associated OAM are crucial for a variety of applications ranging from trapping and manipulation[22,23], quantum entanglement[2], and optical communications[17,18]. Despite a plethora of techniques proposed and demonstrated for OVB generation and vortex detection[11,24], it has always been a challenge to selectively confine and transport even a low-order (e.g., with a topological charge $l = 1$ or $-1$) optical vortex. In a waveguide or a laser cavity, any perturbation that breaks the circular symmetry can lead to vortex deformation. Even in homogeneous linear media, a high-order multiple-charge ($l > 1$) OVB commonly disintegrates into several "pieces" of single-charge vortices during propagation, susceptible to ambient perturbations[25]. Vortex solitons, on the other hand, demand nonlinear media and controlled stability[26]. In fact, currently, there is no effective way to preserve a vortex as a single donut-shaped wave packet traveling from one place to another.

Recently, there has been a surge of interest in the study of topological disclinations[8,9,27-31], unveiling nontrivial topological phases, especially with respect to higher-order topological phases[14-16]. Topological disclinations, a representative type of topological defects of a point-group rotational symmetry, can support localized topological states within the bulk[8,9,32] rather than just at the boundaries[33]. However, lattice defects and disclinations typically break the chiral symmetry (CS) - a holy grail element for topological protection in a large family of topological insulators[12]. If disclinations are designed to preserve the lattice CS, their bounded states can lie at mid-gap and be pinned at zero-energy[30]. Such chiral symmetric disclinations are ideal for the realization of a topological vortex guide (TVG), where spectral isolation, spatial confinement, and topological protection can be guaranteed. Nevertheless, a single-channel TVG has never been realized, for which one would need a disclination structure with a single-site core that meanwhile respects the CS, and one would also need to consider complex vortex (high-orbital) mode coupling over the entire lattice structure. Indeed, most topological disclination structures established so far do not possess chiral symmetry, and there is no mid-gap state at the disclination core that typically consists of several lattice sites[8,9]. Apart from the $p$-orbital disclination states[34], topological vortex mid-gap states have never been realized. In recent endeavors of localized vortex states with a topological defect[35], the bounded vortex modes belong to the lowest $s$-orbital modes only, occupying multiple sites in a discrete lattice. It is important to point out that, in a Dirac-vortex topological cavity or lattice, the "vortex" refers to a Kekulé modulation of a Dirac lattice with a vortex phase (as in the Jackiw-Rossi model)[36-38], but the bounded propagating mode itself is not a vortex of high orbitals.

In this work, we demonstrate a universal principle for realization of perfect TVGs by use of specially designed topological disclination structure. As illustrated in Fig. 1a, a vortex beam preserves its donut shape and phase singularity along the central waveguide channel – the single-site disclination core of a $C_3$-rotational and chiral symmetric lattice. It is *doubly* protected by both nontrivial momentum-space ($k$-space) band topology and real-space topology characterized by the nontrivial winding of the complex vortex-mode coupling (Fig. 1b). The $k$-space band topology gives rise to the localized mid-gap vortex states protected by the CS, while the real-space nontrivial winding ensures that we can selectively excite and transport just a single vortex mode in the central waveguide (without interfered by all other modes that have opposite-vorticity components at the core). Although illustrated with a $C_3$-symmetric disclination, the principle applies to high-order vortices with topological charge $l$ for any rotational order $n$ of the lattice symmetry $C_n$, as long as $2l/n$ is not an integer. This condition is well characterized as the vorticity-coordinated rotational symmetry (VRS) – unique for the protection of a single vortex transport in topological disclinations. Experimentally, we employ laser-written photonic lattices and show that both $l = 1$ and $l = 2$ vortices are guided by a nontrivial $C_3$-symmetric disclination structure, manifesting a zero-energy mode in the mid-gap. In contrast, in a $C_4$-symmetric structure the $l = 2$ vortex breaks up into a quadrupole-like pattern due to disobeying of the VRS. Numerical simulations (Fig. 1c) further corroborate the necessary protections for vortex guidance. Our work represents the first demonstration of single-channel TVGs for robust transport of vortices in any system[39].

**Experimental Results:**

Our disclination lattices are constructed by a modified 'cutting and gluing' procedure[8,9] that results in a single-site core in the center as a topological defect (see Methods; Extended Data Fig. 1). Experimentally, we establish two types of such lattices characterized by $C_3$- and $C_4$-rotational symmetry using a CW-laser-writing technique (Supplementary note 1). A typical example of the $C_3$ disclination and corresponding results are presented in Fig. 2, where Fig. 2a1 is the disclination lattice and Fig. 2a2 is the input OVB used to probe the central disclination core. The input size of the OVB is $36\ \mu$m, and it expands to about $220\ \mu$m after 20-mm propagation in free space, as seen from the interferogram (Fig. 2b1) which exhibits a single bifurcation of fringes (indicating $l = 1$)[11,25]. In contrast, when the same OVB is launched into the disclination core, its intensity is well confined in the core, preserving both the donut-shaped pattern and the topological charge (Fig. 2b2). The 3D intensity plots in Fig. 2 illustrate clearly the difference between a guided and an unguided vortex. Results from numerical simulation (Fig. 2b3) agree well with the observation, showing the robustness of the guided vortex even at much longer propagation distances (Methods; Extended Data Fig. 2; and Supplementary note 2). As seen in Fig. 2c1-c3, a high-order OVB with $l = 2$ is also guided in the disclination core of the $C_3$-lattice, although it expends to a larger ring pattern in free space (Fig. 2c1). However, the $l = 2$ vortex remains intact during propagation in the $C_3$-lattice, but it disintegrates into a quadrupole-like pattern in the $C_4$-lattice (Fig. 1c), as observed in our experiment.

**Disclination vortex states from momentum-space topology**

To understand the essence of double protection needed for the TVGs, let us start from the $k$-space band topology, characterized by the theory of topological invariants in momentum space[12,40]. The band structure of the disclination Hamiltonian considering complex vortex-mode coupling for a $C_n$-symmetric lattices are calculated using the tight-binding model (see Methods and Supplementary note 3). In our model, the $k$-space band topology lies in a chiral symmetric topological phase, which ensures that the topological defect states appear right at zero-energy in the mid-gap (see Fig. 3a1-a2 for $l = 1$ and $l = 2$ cases in the $C_3$ lattice) and occupy only one sublattice (Fig. 3b1-b2; 3c1-c2). Because the number of sites in every unit cell is not the same, such defect states cannot be characterized by the topological invariants conventionally used for disclination structures, e.g., the fractional charge density[8,15,29]. To solve this issue, we employ the concept of multipole chiral number (MCN)[16] - a bulk integer topological invariant $N$ developed recently for predicting higher-order topological phases of chiral-symmetric structures without the need for periodic boundary conditions. We generalize the MCN concept so it can be applied to the $C_n$-symmetric disclination lattices even when the number of lattice sites belonging to different sublattices are not equal (Fig. 1a) (Supplementary note 4). Here, the MCN evaluates the difference in multipole momentum that originated from A and B sublattices in the $C_3$ lattice (Fig. 3d-3e), and it physically describes the winding of the phase term of the wave function on the B sublattice with respect to the A sublattice. For the whole structure of a $C_n$ lattice, we use coordinates distinctively for $n$ sectors, so as to define multipole operators with respect to the centre defect to get $N$. The system is topologically nontrivial when $d_2 < d_1$, where $d_1$ and $d_2$ are the waveguide distances relating to the intracell and intercell coupling, respectively. In this case, a nonzero $N$ corresponds to nontrivial winding in momentum space, as pictured in Fig. 1b. For example, in Fig. 3d, $N = 2$ indicates there are two degenerate zero-energy vortex states (Fig. 3a1, 3a2) with an opposite circulation of the vortex phase. In contrast, if $d_2 > d_1$, we have $N = 0$ (Fig. 3d), implying a topologically trivial winding without topological defect states. Results for some other $C_n$ lattices are presented in Extended Data Fig. 3. This generalized MCN can be applied to characterize higher-order topological phases in other non-periodic $C_n$-symmetric structures with chiral symmetry.

**Selection and protection of a single vortex from real-space topology**

Notwithstanding that there are two degenerate zero-energy vortex modes, robust transport of a vortex requires that only a single vortex state should be preserved during propagation. There are two types of vortex mode couplings between waveguides that can support both clockwise ($l < 0$) and anti-clockwise ($l > 0$) phase circulations. As illustrated in Fig. 4a1, the same-vorticity mode coupling (SVMC) between two vortex modes is always real for couplings from any direction, but opposite-vorticity mode coupling (OVMC) is direction-dependent, which is better pictured by a coupling vector $t_{OV}$ in the complex plane (Fig. 4a2). To ensure that a TVG supports only a single clockwise (or anti-clockwise) vortex mode at any propagation distance, no anti-clockwise (or clockwise) components should intervene in the transport, as analyzed in Supplementary note 5. This requires that the collective contribution of OVMC from all waveguides to the disclination core must be zero.

In a $C_n$ disclination lattice, we evaluate the OVMC between the core and all other waveguides belonging to each distinct sector, and then examine the winding of the complex coupling $T_j (j = 1,2, ... n)$ from each of the $n$ rotational sectors (see Fig. 4b1-b3 for $C_3$ lattice and Fig. 4c1-c3 for $C_4$ lattice). Real-space topology can be unveiled by defining a coupling winding number as

$$w = \frac{1}{2\pi i} \sum_{j=1}^{n-1} \ln\left(\frac{T_{j+1}}{T_j}\right) \qquad (1)$$

We find that $w$ is nonzero only when $2l/n$ is not an integer number, indicating the existence of a topologically nontrivial phase, and in this case the total complex coupling $\sum_j^n T_j = 0$. On the contrary, where $2l/n$ is an integer, $w$ becomes zero, and also $\sum_j^n T_j \neq 0$ (Supplementary note 6). In the example of $C_3$ lattice, we find that the coupling winding is non-zero for both $l = 1$ and $l = 2$ (Fig. 4b2, b3), thus resulting in protected vortex transport (see Fig. 2). However, in the $C_4$ lattice (Fig. 4d1), the winding is non-zero for $l = 1$ but vanishes for $l = 2$. As such, the $C_4$ lattice supports a single-charge vortex (Fig. 4d2) but not a double-charge vortex. In this latter case, the $l = 2$ vortex breaks up into a quadrupole-like pattern (Fig. 4d3), in accordance with simulation results (see Fig. 1c and Extended Data Fig. 2). This winding picture (see also Fig. 1b) resembles a skyrmion-like spin texture in a magnetic structure. Here, nontrivial real-space winding of the coupling vectors needs to be "coordinated" by the rotational order of symmetry $n$ and the topological charge $l$ of the guided vortex – the reason why we name it the "VRS".

We further explore the general cases with arbitrary $n$ and $l$, and theoretically prove that the total complex coupling accounting for the OVMC on the complex plane between the central defect waveguide and all contributions originating from $n$ sectors cancels if and only if the topological invariant $w$ is nonzero, which requires $2l/n$ is not an integer (Supplementary notes 5-6). Such a condition (summarized in Fig. 4a3) sets up a universal rule for protecting the transport of only a single high-order vortex mode (clockwise or anti-clockwise, but not both) along a single channel in the $C_n$ lattice, as if all other waveguides with OVMC were absent (Supplementary note 7). Take an alternative example of a $C_5$ disclination, we show that an OVB with $l = 5$ or $l = 10$ cannot maintain its shape during propagation, but other high-order vortices are well guided in the disclination core so long a noninteger $2l/5$ is guaranteed (Extended Data Fig. 4).

**Discussion and outlook**

Vortices in superconductors and superfluids are intriguing examples of topological defects, just as skyrmions, domain walls, cosmic strings, and monopoles that have intrigued scientists in diverse fields over the decades. While those topological defects can be considered as quasiparticles with some stability against perturbations, a vortex wave packet such as an OVB cannot self-localize by itself. We have demonstrated a fundamental principle for confinement and transport of vortices by the use of chiral-symmetric topological disclinations. One may wonder why not to use a single circular waveguide (such as an optical fiber) in lieu of the disclination structure. Briefly, a single waveguide cannot provide topological protection. Even in a specially engineered few-mode "vortex" fiber, the degenerate pair of OAM states inevitably mix into each other while the chosen OAM mode also leaks into other parasitic modes after long propagation[17]. In a conventional single-core fiber, apart from unwanted mode coupling, a chosen OAM mode is also afflicted with ambient perturbation during transport. In Supplementary note 8, we compare the vortex beam propagation under perturbation in a single waveguide and in a $C_3$ disclination structure, illustrating that an OVB cannot preserve its structure during transport in a single waveguide even with perturbations respecting rotational symmetry, while it remains intact in the disclination structure due to topological protection. Detailed stability analysis under different symmetry-preserving perturbations (such as CS, sub-symmetry[41], and rotational symmetry) are presented in Supplementary notes 9-10, confirming the necessity of double protection for TVGs.

Our work, although demonstrated in a photonic platform, may bring about a solution to the long-standing challenge in the control of vortex waves or swirling flows, as the underlying physics for topological protection of vortices is broadly valid for diverse systems. It can be readily applied to different fields where controlled vortex transport is desired, including hydrodynamics, acoustics, aeronautics, heat exchange, and combustion. In photonics, our work may herald new photonic devices in addition to TVGs: for example, it may be readily adopted for the design of microstructured fibers for protected OAM modes to enable uncorrupted information networks[17,18,42] and for the realization of ultra-compact vortex lasers and topological cavities[19,36].

**Methods:**

**Construction of single-site disclination lattices**

The $C_n$-symmetric disclination used in this work is constructed by a modified "cutting and gluing" procedure[8,9]. Compared to previous models derived from standard two-dimensional (2D) Su-Schrieffer-Heeger (SSH) models[43], our lattice structure constitutes a design for a single-core chiral symmetric disclination.

As illustrated in Extended Data Fig. 1a, b, the conventional disclination structures belong to either type-I or type-II categories[8,9]. The type of disclination is identified by the amount of translation and rotation of a vector around the path itself (i.e., translation value [$a$] and Frank angle $\Omega$)[29]. Type-I disclination lattices with Frank angle $\Omega = -90°$ and the holonomy value of a closed path around the core $[a]^{(4)} = 0$ are terminated by weak bonds at the center location [Extended Data Fig. 1a]. Complementarily, type-II lattices with $\Omega = -90°$, $[a]^{(4)} = 1$ have strong bonds around the defect core [Extended Data Fig. 1b]. In the nontrivial phase, the Wannier centers (yellow quadrangular stars) are positioned at the intersection among four-unit cells. We note that chiral symmetry is not preserved in both types of disclinations[8].

The $C_n$-disclination with a single-site core used in our work cannot be categorized as one of the above classes, and its formation comports the removal of some lattice sites instead of just "cutting and gluing". To guarantee the existence of zero-energy bound states, we appropriately modify an initial type-II disclination structure. It displays three-unit cells composed of four sites intersecting at the center, each of which belongs to one of the $C_3$-symmetric sectors (see Extended Data Fig. 1c). We first shrink every lattice sector with respect to the core until the three nearest waveguides overlap perfectly. The white arrows in the inset indicate the direction of shrinking. Then, any overlapped (extra) waveguides which break the CS in the traditional disclination structure are removed, so the array index is still uniform. In a similar way, other $C_n$-symmetric lattices can be readily constructed. A characteristic marking the difference from previous type-I and -II disclination models is that now the single-core $C_n$-disclinations possess chiral symmetry (no coupling between the same sublattice sites), so they can support topologically protected zero-energy bound states.

**Discrete Vortex Hamiltonian**

In the OAM domain, we can express the real-space Hamiltonian of a $C_n$-symmetric disclination lattice with topological charge $l$ under the tight-binding approximation as

$$H = \sum_{\mathbf{R},\mathbf{R}',l} \xi(\mathbf{R} - \mathbf{R}') \left[ \kappa_{\text{SV}} c^{\dagger}_{\mathbf{R},l} c_{\mathbf{R}',l} + \kappa_{\text{OV}} e^{2il\theta(\mathbf{R}-\mathbf{R}')} c^{\dagger}_{\mathbf{R},l} c_{\mathbf{R}',-l} \right]. \tag{M1}$$

where $\xi(\mathbf{R} - \mathbf{R}') = e^{-\rho|\mathbf{R}-\mathbf{R}'|}$ is the hopping amplitude between two nearest-neighbor waveguides of $C_n$-disclination lattice located at the positions $\mathbf{R}$ and $\mathbf{R}'$, and $\rho$ is a scale factor. The hopping amplitudes are approximated as an exponential decay function of the spacing $|\mathbf{R} - \mathbf{R}'|$[44] . The coefficient $\kappa_{SV}$ describes the SVMC while $\kappa_{OV}$ the OVMC, with $\theta(\mathbf{R} - \mathbf{R}')$ being the azimuth angle of the vector $\mathbf{R} - \mathbf{R}'$. $c_{\mathbf{R},l}^\dagger$ is the creation operator at the lattice site with position $\mathbf{R}$ corresponding to a vortex mode with topological charge $l$. A similar definition is made for the other creation and annihilation operators. The vortex band structures reported in Fig. 3 are calculated by diagonalizing $H$ for the same $C_3$-disclination structure but distinct $l$ values. Related vortex-mode distributions are found by retrieving both anti-clockwise ($l > 0$) and clockwise ($l < 0$) components from calculated eigenvectors of $H$.

**Experimental Methods**

In the experiment, two disclination photonic lattices with $C_3$- and $C_4$-rotational symmetries are established in a 20 mm-long nonlinear crystal by employing a site-to-site CW-laser-writing technique[45,46]. An ordinary-polarized laser beam with 532 nm wavelength and low power is phase modulated by a spatial light modulator (SLM) in the Fourier domain, in order to create a quasi-nondiffracting beam at variable writing positions. Every waveguide remains intact during each set of measurements due to the photorefractive memory effect[45,46]. The probing process is performed by launching an extraordinary-polarized vortex beam with either single or double topological charges as created by the SLM, injected into the disclination core. Finally, interferograms are obtained from interference between the probe beam and a quasi-plane wave (see Supplementary note 1 for more details).

**Numerical Methods**

The propagation dynamics of an optical vortex beam are simulated by the continuum model of a nonlinear Schrödinger-like equation (NLSE) [42, 41]

$$i\frac{\partial \Psi}{\partial z} = -\frac{1}{2k}\nabla_\perp^2 \Psi - \frac{k\Delta n_0}{n_0}\frac{\Psi}{1 + I_L(x,y) + I_P(x,y)}, \quad (M2)$$

where $\Psi(x, y, z)$ is the electric field envelope, with $x$ and $y$ denoting the transverse coordinates, $z$ is the propagation distance, and $\nabla_\perp^2 = \partial^2/\partial x^2 + \partial^2/\partial y^2$ is the transversal Laplacian operator. $k$ is the wavenumber in the medium, $n_0 = 2.35$ is the refractive index for the specific photorefractive crystal and $\Delta n_0 = -n_0^3 r_{33} E_0/2$ is the refractive-index change, where $r_{33} = 280$ pm/V is the electro-optic coefficient along the crystalline $c$-axis, and $E_0$ is the bias electric field. The two terms $I_L(x, y)$ and $I_P(x, y)$ denote respectively the intensity patterns of the disclination lattice-writing and the probe beams. To confirm the theoretical prediction of TVG formation in the proposed $C_n$-disclination lattice, measurements from experimental observations are also corroborated with numerical simulations using the NLSE in Eq. (1) (Supplementary note 2). For any initial vortex-beam excitation, the NLSE solutions are found via the split-step Fourier transform method, under the condition that $I_P(x, y, z)$ is weak so the probe beam itself will not undergo nonlinear self-action during propagation.

**Acknowledgments**

We thank Y. Kartashov, Y. Hu and L.M Song for the discussion. This research is supported by the National Key R&D Program of China under Grant No. 2022YFA1404800, the National Natural Science Foundation (12134006, 11922408), and the QuantiXLie Center of Excellence, a project co-financed by the Croatian Government and European Union through the European Regional Development Fund - the Competitiveness and Cohesion Operational Programme (Grant KK.01.1.1.01.0004). D.B. acknowledges support from the 66 Postdoctoral Science Grant of China and the National Natural Science Foundation (12250410236). R.M. acknowledges support from NSERC and the CRC program in Canada.

**Conflict of interests**

The authors declare no conflicts of interest. The authors declare no competing financial interests.

**Contributions**

All authors contributed to this work.

Correspondence and requests for materials should be addressed to Z.C. or H.B.

Supplementary information accompanying the manuscript is available on the website.

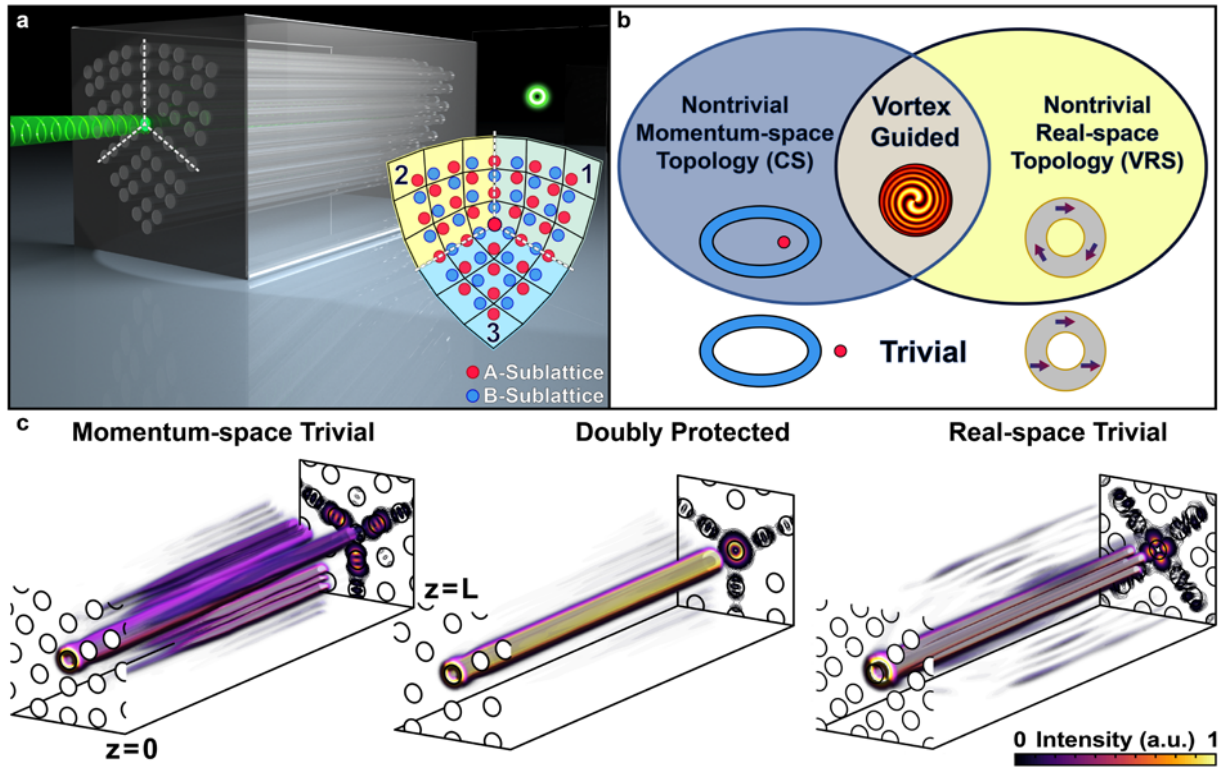

**Fig. 1 Illustration and simulation of robust vortex transport by topological disclinations**. **a** Schematic of a vortex traveling through the core of a disclination lattice which has $C_3$ rotational symmetry and chiral symmetry (CS) with two sublattices and a single-site core as shown in the inset. **b** Venn diagram of the underlying topology associated with vortex guidance. The light blue region represents nontrivial momentum-space winding (illustrated with a nonempty winding loop) as for a CS structure. The yellow region represents nontrivial real-space winding (illustrated with winding complex coupling vectors akin to skyrmion-like topological winding in a magnetic structure) when the disclination lattice has the vorticity-coordinated rotational symmetry (VRS) discussed in the text. The overlapping region represents a "doubly protected" topologically nontrivial structure which can lead to the topological vortex guide (TVG). All other regions are topologically trivial. **c** Numerical simulations showing confined vortex transport through a TVG in the doubly protected structure, while the high-order vortex beam expands and breaks up in an unprotected structure, corresponding to the three colored regions in **b**.

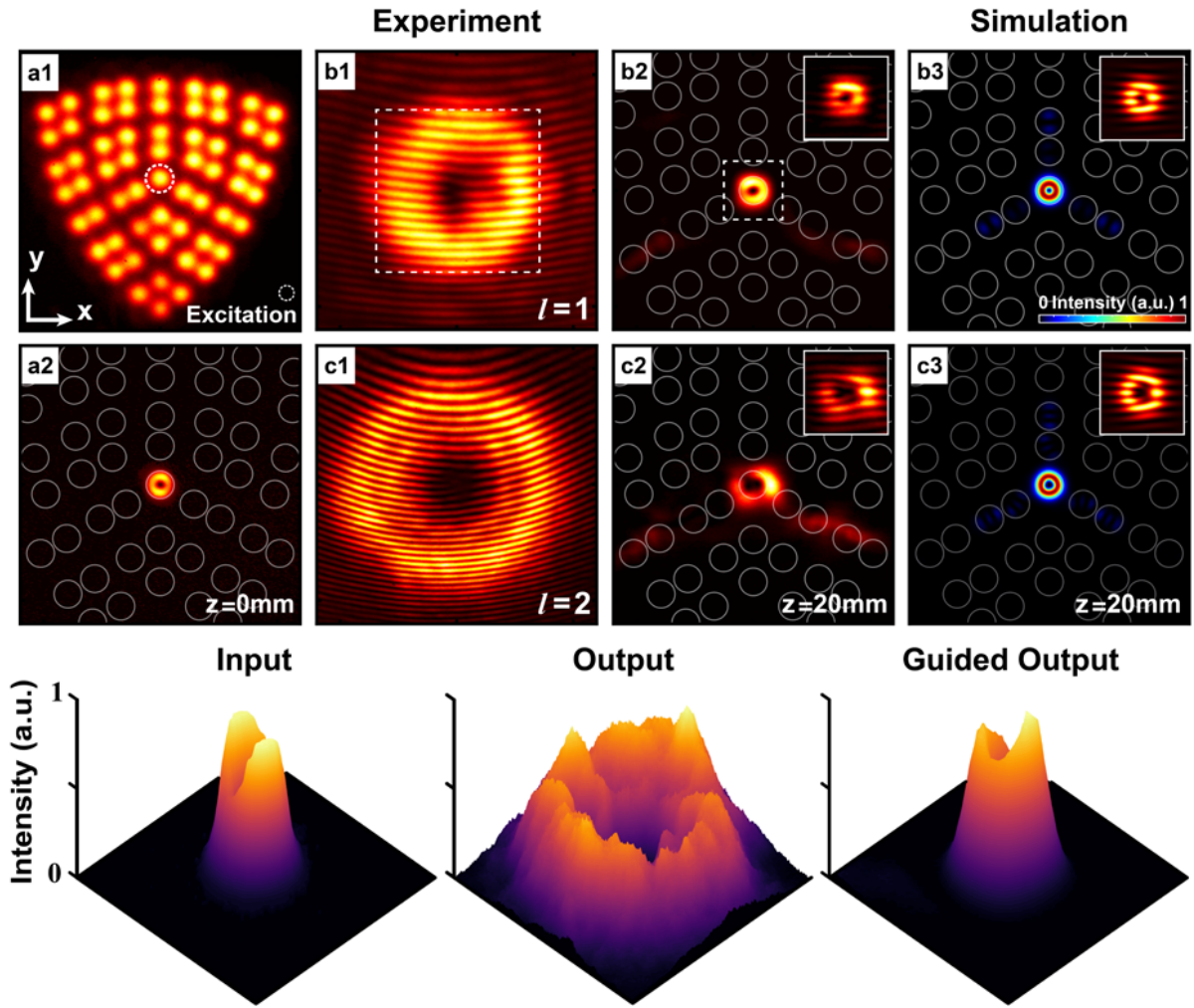

**Fig. 2 Experimental demonstration of a TVG by chiral-symmetric disclinations.** (a1) A laser-written photonic lattice of $C_3$ rotational symmetry and a single-site disclination core, where a white dashed circle marks the core location for excitation with a vortex beam. (a2) Input intensity pattern of the vortex beam probing the TVG. (b1) Output interferogram of a single-charge $l = 1$ vortex beam after 20-mm free propagation without the lattice. (b2, b3) Output intensity pattern of the vortex beam after propagating through the disclination lattice from (b2) measurement and (b3) continuum-model simulation. Top-right insets are corresponding interferograms confirming the preserved vortex phase. (c1-c3) Corresponding results obtained for a high-order vortex probe with double-charge $l = 2$. Bottom panels are 3D intensity plots of experimental results from the $l = 1$ vortex beam.

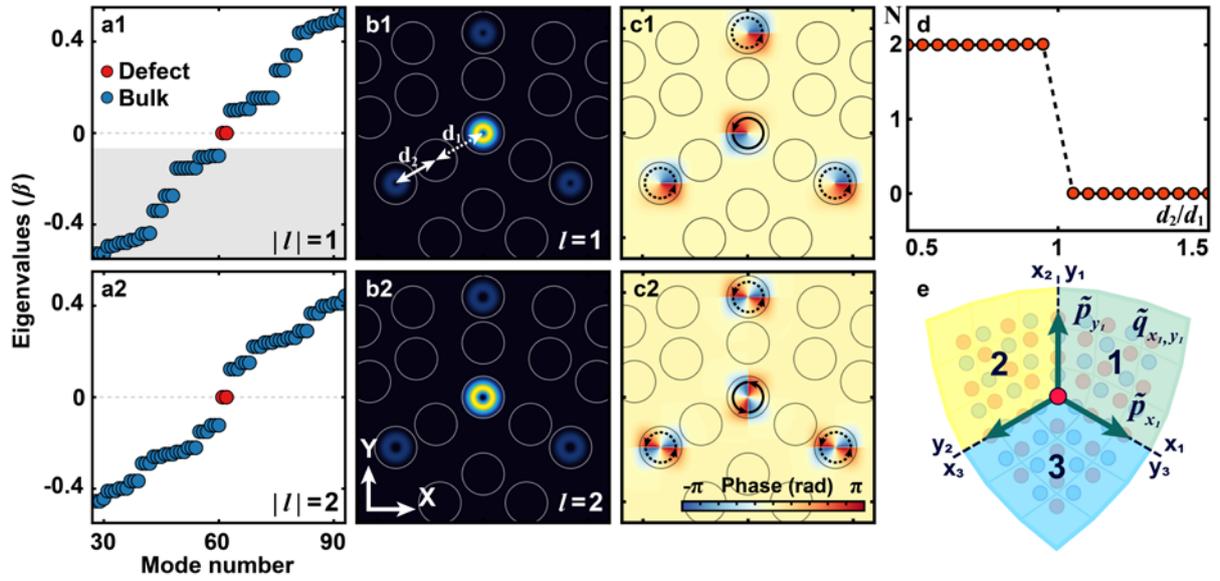

**Fig. 3 Zero-energy disclination vortex modes protected by band topology. a1** Calculated eigenvalues of single-charge vortex modes in $C_3$ disclination structure, where two degenerate vortex modes (red) appear right at mid-gap but with opposite vorticity. The intensity and phase distribution of one of them ($l = 1$) are plotted in **b1, c1**, showing confinement mostly at the disclination core while exponentially decaying "tails" distribute only in the same (next-nearest-neighbor) sublattice sites with a $\pi$-phase difference – a characteristic of the topological states. **a2-c2** Corresponding results obtained for high-order ($l = 2$) vortex modes. Notice the difference of phase distribution between **c1** and **c2**, but the core and "tails" again have a $\pi$-phase difference in **c2**. (d) Calculated topological invariant - the multipole chiral number $N$, which equals 2 when the intracell spacing $d_1$ is larger than the intercell spacing $d_2$, indicating a topologically non-trivial regime with two disclination modes. (e) Illustration of multipole moments in the $C_3$ disclination structure, where $\tilde{q}$ and $\tilde{p}$ are the difference of dipole and quadrupole moments between sublattices, respectively. Here we choose three sets of coordinates corresponding to the three sectors to generate the multipole operators. $N$ is equivalent to the overall difference between the multipole moments of two sublattice wave functions.

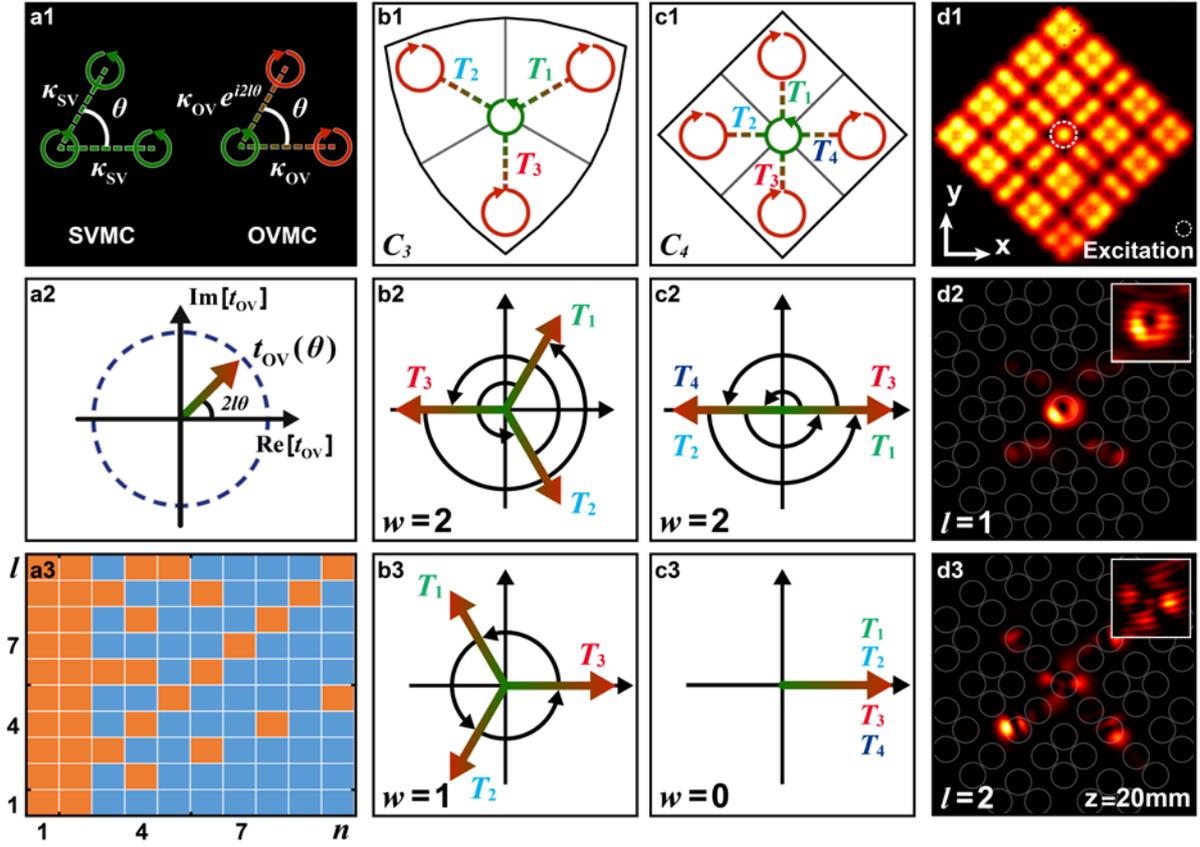

**Fig. 4 Protection of high-order vortices by vorticity-coordinated rotational symmetry (VRS) and nontrivial real-space winding. a1** Illustration of vortex mode coupling between two waveguides where $\kappa$ is the coupling amplitude. The same-vorticity mode coupling (SVMC) is not direction-dependent, but the opposite-vorticity mode coupling (OVMC) is as plotted in **a2**, where the coupling coefficient $t_{OV}$ depends on $\theta$. In a $C_n$ symmetric disclination structure, all coupling contributions to the centre vortex mode can be calculated by sectors as illustrated in **b1** (2$^{nd}$ column) for $C_3$ and **c1** (3$^{rd}$ column) for $C_4$ disclinations, where $T_j$ is the equivalent coupling for all OVMCs in each sector. To guarantee only a single vortex mode ($l$ or $-l$) in the disclination core, $T_j$ must have nontrivial coupling winding number ($w \neq 0$) as shown in **b2, c2** (for $l = 1$) and **b3** (for $l = 2$). This is more suitably described by the VRS which demands a noninteger value of $2l/n$ when the disclination has $n$-fold rotational symmetry; this is summarized in **a3** where orange (blue) indicates unprotected (protected) vortex modes. Take $l = 2$ as an example, it is protected in the $C_3$ (due to nontrivial winding $w = 1$) but unprotected in the $C_4$ disclination. Single vortex mode is preserved only under the non-zero topology winding condition. In the 4$^{th}$ column, experimental results compare a protected $l = 1$ vortex (**d2**) and an unprotected $l = 2$ vortex (**d3**) in $C_4$ disclination, in agreement with the plot of (**a3**).

**References:**

1. Matthews, M. R. *et al.* Vortices in a Bose-Einstein Condensate. *Phys. Rev. Lett.* **83**, 2498-2501 (1999).

2. Mair, A. *et al.* Entanglement of the Orbital Angular Momentum States of Photons. *Nature* **412**, 313-316 (2001).

3. Zwierlein, M. W. *et al.* Vortices and Superfluidity in a Strongly Interacting Fermi Gas. *Nature* **435**, 1047-1051 (2005).

4. Uchida, M. & Tonomura, A. Generation of Electron Beams Carrying Orbital Angular Momentum. *Nature* **464**, 737-739 (2010).

5. Clark, C. W. *et al.* Controlling Neutron Orbital Angular Momentum. *Nature* **525**, 504-506 (2015).

6. Luski, A. *et al.* Vortex Beams of Atoms and Molecules. *Science* **373**, 1105-1109 (2021).

7. Wioland, H. *et al.* Confinement Stabilizes a Bacterial Suspension into a Spiral Vortex. *Phys. Rev. Lett.* **110**, 268102 (2013).

8. Peterson, C. W. *et al.* Trapped Fractional Charges at Bulk Defects in Topological Insulators. *Nature* **589**, 376-380 (2021).

9. Liu, Y. *et al.* Bulk-Disclination Correspondence in Topological Crystalline Insulators. *Nature* **589**, 381-385 (2021).

10. Allen, L. *et al.* Orbital Angular Momentum of Light and the Transformation of Laguerre-Gaussian Laser Modes. *Phys. Rev. A* **45**, 8185-8189 (1992).

11. Shen, Y. *et al.* Optical Vortices 30 Years On: OAM Manipulation from Topological Charge to Multiple Singularities. *Light Sci. Appl.* **8**, 90 (2019).

12. Ozawa, T. *et al.* Topological Photonics. *Rev. Mod. Phys.* **91**, 015006 (2019).

13. Chiu, C.-K. *et al.* Classification of Topological Quantum Matter with Symmetries. *Rev. Mod. Phys.* **88**, 035005 (2016).

14. Benalcazar, W. A., Bernevig, B. A. & Hughes, T. L. Quantized Electric Multipole Insulators. *Science* **357**, 61-66 (2017).

15. Peterson, C. W. *et al.* A Fractional Corner Anomaly Reveals Higher-Order Topology. *Science* **368**, 1114-1118 (2020).

16. Benalcazar, W. A. & Cerjan, A. Chiral-Symmetric Higher-Order Topological Phases of Matter. *Phys. Rev. Lett.* **128**, 127601 (2022).

17. Bozinovic, N. *et al.* Terabit-Scale Orbital Angular Momentum Mode Division Multiplexing in Fibers. *Science* **340**, 1545-1548 (2013).

18. Willner, A. E. *et al.* Optical Communications Using Orbital Angular Momentum Beams. *Adv. Opt. Photon.* **7** (2015).

19. Miao, P. *et al.* Orbital Angular Momentum Microlaser. *Science* **353**, 464-467 (2016).

20. Nye, J. F. & Berry, M. V. Dislocations in Wave Trains. *Proc. R. Soc. Lond. A* **336**, 165-190 (1974).

21. Veshchunov, I. S. *et al.* Optical Manipulation of Single Flux Quanta. *Nat. Commun.* **7**, 12801 (2016).

22. He, H. *et al.* Direct Observation of Transfer of Angular Momentum to Absorptive Particles from a Laser Beam with a Phase Singularity. *Phys. Rev. Lett.* **75**, 826-829 (1995).

23. Paterson, L. *et al.* Controlled Rotation of Optically Trapped Microscopic Particles. *Science* **292**, 912-914 (2001).

24. Ni, J. *et al.* Multidimensional Phase Singularities in Nanophotonics. *Science* **374**, eabj0039 (2021).

25. Soskin, M. S. *et al.* Topological Charge and Angular Momentum of Light Beams Carrying Optical Vortices. *Phys. Rev. A* **56**, 4064-4075 (1997).

26. Desyatnikov, A. S., Kivshar, Y. S. & Torner, L. Optical Vortices and Vortex Solitons. *Prog. Opt.* **47**, 291-391 (2005).

27. Teo, J. C. & Hughes, T. L. Existence of Majorana-Fermion Bound States on Disclinations and the Classification of Topological Crystalline Superconductors in Two Dimensions. *Phys. Rev. Lett.* **111**, 047006 (2013).

28. Wang, Q. *et al.* Observation of Protected Photonic Edge States Induced by Real-Space Topological Lattice Defects. *Phys. Rev. Lett.* **124**, 243602 (2020).

29. Li, T. *et al.* Fractional Disclination Charge in Two-Dimensional Cn-Symmetric Topological Crystalline Insulators. *Phys. Rev. B* **101**, 115115 (2020).

30. Deng, Y. *et al.* Observation of Degenerate Zero-Energy Topological States at Disclinations in an Acoustic Lattice. *Phys. Rev. Lett.* **128**, 174301 (2022).

31. Lin, Z.-K. *et al.* Topological Phenomena at Topological Defects. arXiv:2208.05082v05081 [cond-mat.mtrl-sci] (2023).

32. Lustig, E. *et al.* Photonic Topological Insulator Induced by a Dislocation in Three Dimensions. *Nature* **609**, 931-935 (2022).

33. Noh, J. *et al.* Topological Protection of Photonic Mid-Gap Defect Modes. *Nat. Photon.* **12**, 408-415 (2018).

34. Chen, Y. *et al.* Observation of Topological P-Orbital Disclination States in Non-Euclidean Acoustic Metamaterials. *Phys. Rev. Lett.* **129**, 154301 (2022).

35. Wang, Q. *et al.* Vortex States in an Acoustic Weyl Crystal with a Topological Lattice Defect. *Nat. Commun.* **12**, 3654 (2021).

36. Gao, X. *et al.* Dirac-Vortex Topological Cavities. *Nat. Nanotechnol.* **15**, 1012-1018 (2020).

37. Menssen, A. J. *et al.* Photonic Topological Mode Bound to a Vortex. *Phys. Rev. Lett.* **125**, 117401 (2020).

38. Ma, J. *et al.* Nanomechanical Topological Insulators with an Auxiliary Orbital Degree of Freedom. *Nat. Nanotechnol.* **16**, 576-583 (2021).

39. Hu, Z. *et al.* Topological Guidance of Vortices by Disclination. In Conference on Lasers and Electro-Optics Optics. Paper FM2B.3 (OSA, 2023).

40. Hasan, M. Z. & Kane, C. L. Colloquium: Topological Insulators. *Rev. Mod. Phys.* **82**, 3045-3067 (2010).

41. Wang, Z. *et al.* Sub-Symmetry-Protected Topological States. *Nat. Phys.* (2023). https://doi.org/10.1038/s41567-023-02011-9

42. Wong, G. K. *et al.* Excitation of Orbital Angular Momentum Resonances in Helically Twisted Photonic Crystal Fiber. *Science* **337**, 446-449 (2012).

43. Liu, F. & Wakabayashi, K. Novel Topological Phase with a Zero Berry Curvature. *Phys. Rev. Lett.* **118**, 076803 (2017).

44. Jörg, C. *et al.* Artificial Gauge Field Switching Using Orbital Angular Momentum Modes in Optical Waveguides. *Light Sci. Appl.* **9**, 150 (2020).

45. Xia, S. *et al.* Unconventional Flatband Line States in Photonic Lieb Lattices. *Phys. Rev. Lett.* **121**, 263902 (2018).

46. Hu, Z. *et al.* Nonlinear Control of Photonic Higher-Order Topological Bound States in the Continuum. *Light Sci. Appl.* **10**, 164 (2021).

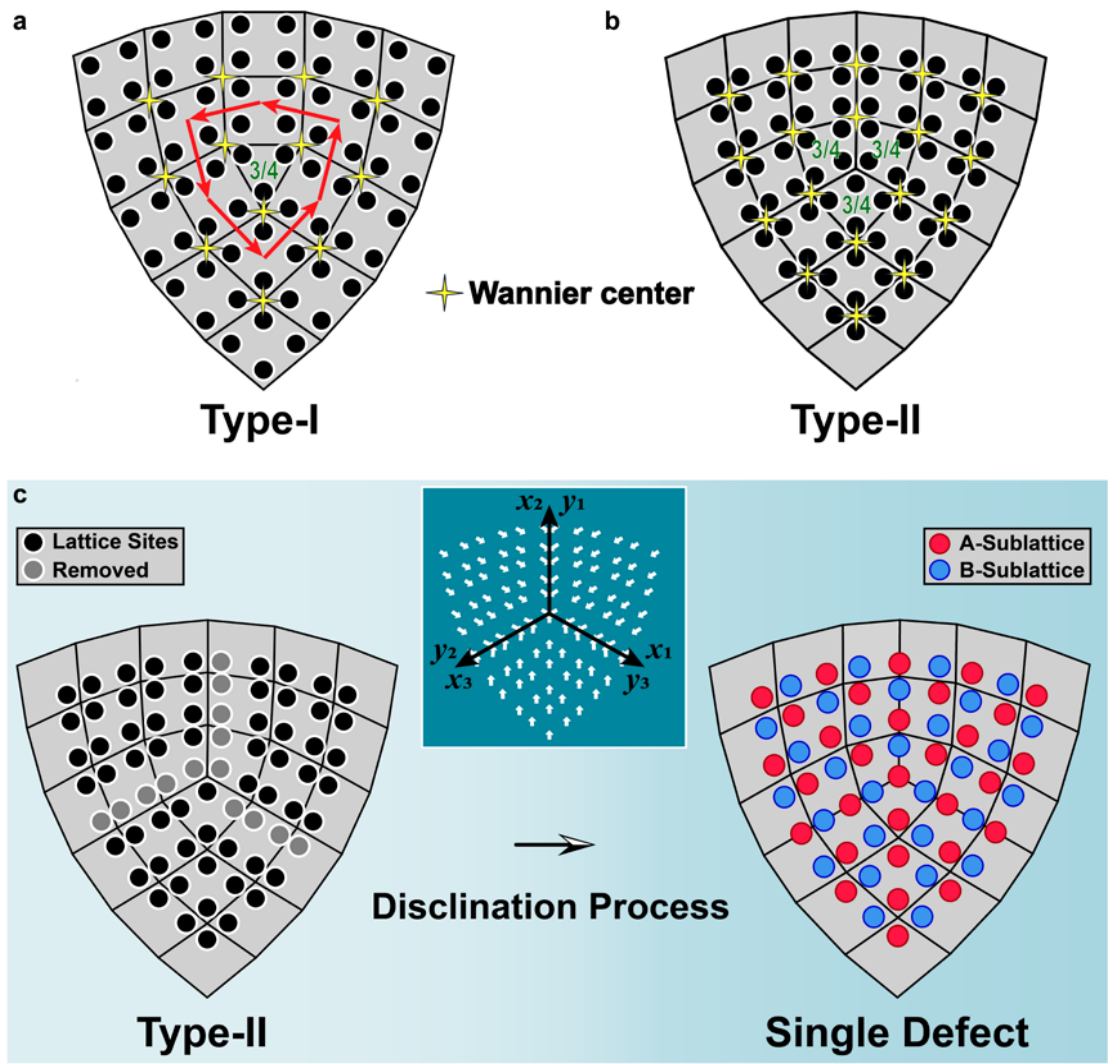

**(Extended Data Fig. 1). Realization of $C_3$-symmetric disclination lattices with a single-site core at the center.** (a, b) Schematics of conventional disclination structures with $C_3$-symmetry, constructed from a standard 2D SSH model through a "cutting and gluing" procedure, belonging to (a) type-I and (b) type-II classifications[29]. Shaded gray quadrilaterals highlight the unit cells, where the yellow stars correspond to the Wannier centers, the black circles are lattice sites, and the red arrows illustrate the calculations of Frank angle $\Omega$ for the type-I lattice[8]. (c) Illustration of the disclination process to realize a single-site-core $C_3$-lattice from the associated type-II disclination structure. Gray circles mark those removed waveguides from the original type-II lattice array, red and blue dots are lattice sites for A and B sublattices, respectively, and the white arrows in the upper middle inset indicate the shifting and merging directions.

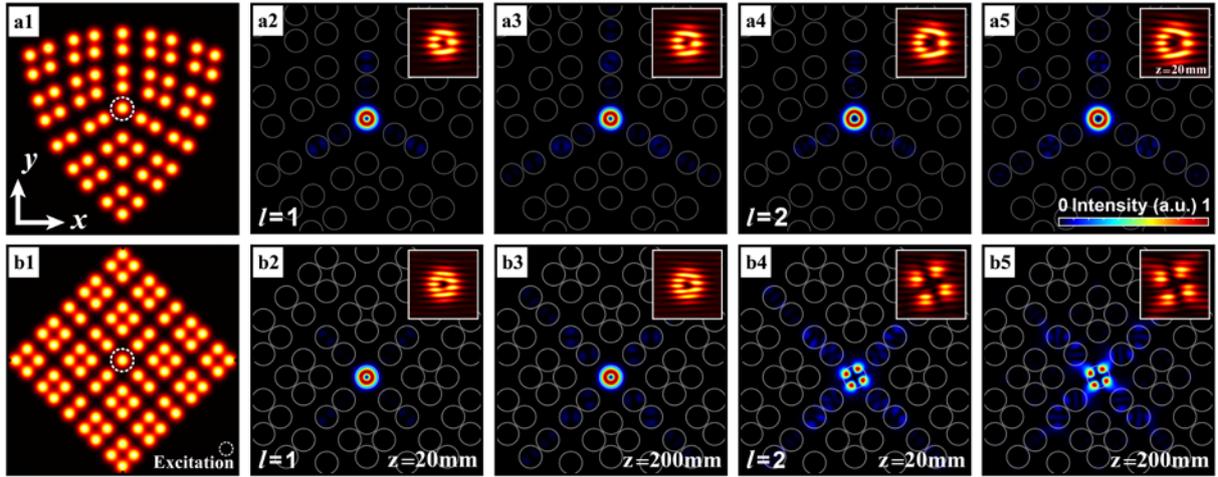

**(Extended Data Fig. 2). Numerical simulations of the topological vortex guide by chiral-symmetric disclinations.** Propagation of the probe vortex is simulated under single-site excitation at the core (circled in a1 and b1) of $C_3$- and $C_4$-symmetric disclination lattices. (a1) The $C_3$ disclination lattice. (a2-a5) Output intensity distributions of a (a2, a3) $l = 1$ and (a4, a5) $l = 2$ vortex beam from the lattice at selected distances (a2, a4) $Z = 20$ mm and (a4, a5) $Z = 200$ mm, highlighting the protected topological vortex guide. Insets are corresponding numerical interferograms, showing the phase singularity from fringe bifurcations. (b1-b5) Same layout as in (a1-a5) except that the results are from the $C_4$-symmetric lattice. In this case, the $l = 1$ vortex is doubly protected by topology, but the $l = 2$ vortex is not protected, broken into a quadruple-like structure at the core with tails populating both sublattices due to the lack of real-space topological protection.

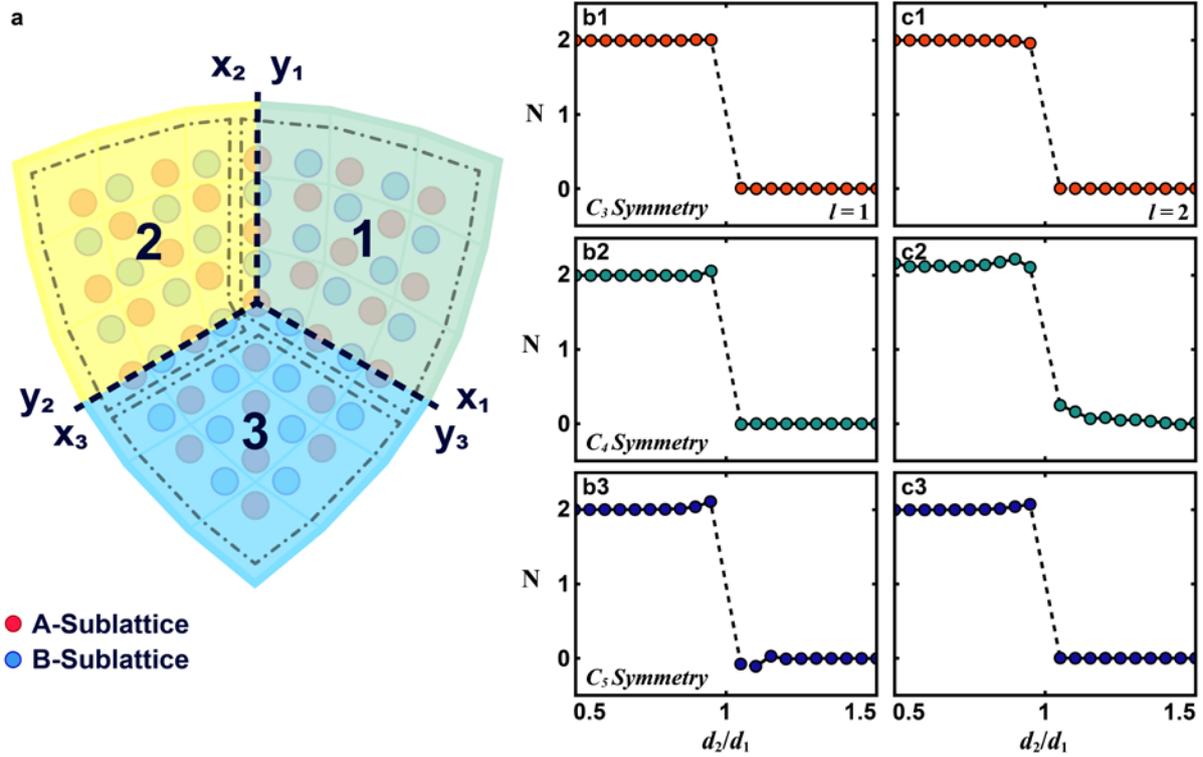

**(Extended Data Fig. 3). Generalized multipole chiral number for describing the topological phase of $C_n$-symmetric disclination lattices.** (a) Schematic illustration showing the definition of coordinate axes in the three groups of a nontrivial $C_3$-symmetric disclination lattice marked by the dash-dotted lines. (b1,c1) Topological invariant of multipole chiral number $N$, calculated for a $C_3$-symmetric disclination lattice for (b1) $l = 1$ and (c1) $l = 2$ vortex modes. (b2, c2) Corresponding calculations for a $C_4$-symmetric lattice, and (b3, c3) for a $C_5$-symmetric lattice.

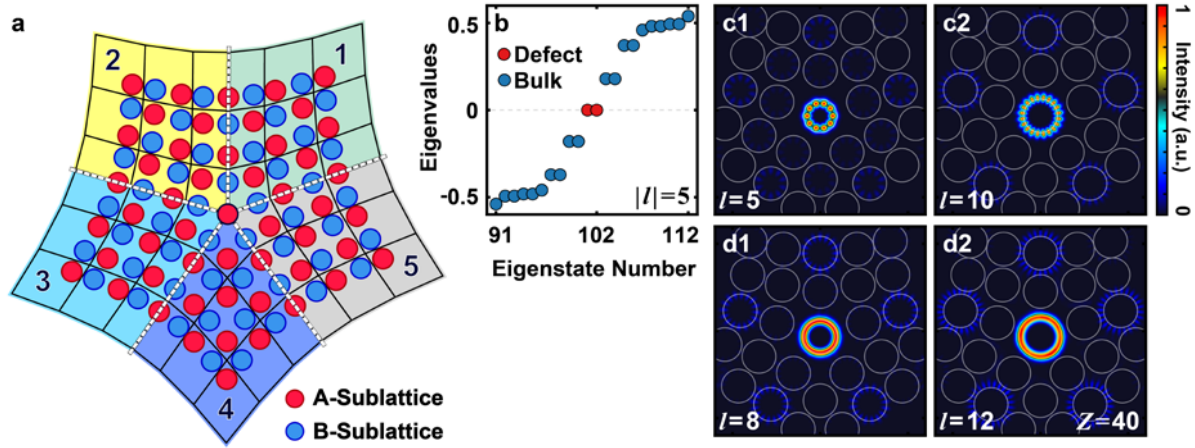

**(Extended Data Fig. 4). Protected and unprotected high-order vortex modes in a $C_5$-symmetric disclination lattice.** (a) Schematic illustration of the chiral-symmetric $C_5$ disclination lattice. (b) Calculated eigenvalues for the $|l| = 5$ vortex modes of the lattice, where two degenerate defect modes appear right at zero energy but with opposite phase vorticity. (c1, c2) Two examples showing *unprotected* high-order vortex modes with topological charge $l = 5$ (c1) and $l = 10$ (c2) after a propagation distance $Z = 40$ (from CMT evolution). In these cases, the VRS is not satisfied because $2l/n$ is an integer. (d1,d2) Two examples showing *protected* high-order vortex modes with topological charge $l = 8$ (d1) and $l = 12$ (d2). In these latter cases, the VRS is satisfied because $2l/n$ is not an integer.